 \documentclass[natbib]{svjour3}                     
\smartqed  
 \usepackage{epsfig}
 \usepackage{mathptmx}      
 \usepackage{latexsym}

\newcommand{\kms}{\hbox{km s$^{-1}$}}
\newcommand{\units}{\hbox{cm$^{-2}$  s$^{-1}$ sr$^{-1}$ keV$^{-1}$}}
\newcommand{\microG}{\hbox{$ \mu{  G}$}}

\newcommand{\OI}{\hbox{O${^\circ}$}}

\newcommand{\nHII}{\hbox{$n({  \mathrm{H}^{+}})$}}
\newcommand{\npro}{\hbox{$n({  \mathrm{p}^{+}})$}}
\newcommand{\nel}{\hbox{$n({  \mathrm{e}^{-}})$}}
\newcommand{\nHI}{\hbox{$n({  \mathrm{H}^\circ})$}}
\newcommand{\HeI}{\hbox{  He$^\circ$}}
\newcommand{\HI}{\hbox{  H$^\circ$}}

\newcommand{\cc}{\hbox{cm$^{-3}$}}
\newcommand{\deeg}{\hbox{$^\circ$}}
\newcommand{\elon}{\hbox{$\lambda$}}
\newcommand{\elat}{\hbox{$\beta$}}

\newcommand\apj{{ApJ}}%
\newcommand\apjl{{ApJ}}%
\newcommand\aap{{A\&A}}%
\newcommand\ssr{{Space~Sci.~Rev.}}%
\newcommand\nat{{Nature}}%
\newcommand\jgr{{J.~Geophys.~Res.}}%
\newcommand\planss{{Planet.~Space~Sci.}}%
\journalname{SSrv}
\begin{document}

\title{The Interstellar Boundary Explorer (IBEX):} \subtitle{Tracing
the Interaction between the Heliosphere and Surrounding Interstellar
Material with Energetic Neutral Atoms}

\titlerunning{IBEX ENA data}        

\author{Priscilla C. Frisch \and David J. McComas
}

\authorrunning{Frisch and McComas} 

\institute{P. C. Frisch\at
              University of Chicago
		Dept. Astronomy and Astrophysics, Chicago, Illinois\\
              Tel.: 773-702-0181\\
              Fax: 773-702-8212\\  
              \email{frisch@oddjob.uchicago.edu}           
           \and
           D. J. McComas \at
              Southwest Research Research Institute and University of
           Texas, San Antonio, Texas
}

\date{Accepted}

\maketitle

\begin{abstract}

The Interstellar Boundary Explorer (IBEX) mission is exploring the
frontiers of the heliosphere where energetic neutral atoms (ENAs) are
formed from charge exchange between interstellar neutral hydrogen
atoms and solar wind ions and pickup ions.  The geography of this
frontier is dominated by an unexpected nearly complete arc of ENA
emission, now known as the IBEX 'Ribbon'.  While there is no consensus
agreement on the Ribbon formation mechanism, it seems certain this
feature is seen for sightlines that are perpendicular to the
interstellar magnetic field as it drapes over the heliosphere.  At the
lowest energies, IBEX also measures the flow of interstellar H, He,
and O atoms through the inner heliosphere.  The asymmetric oxygen
profile suggests that a secondary flow of oxygen is present, such as
would be expected if some fraction of oxygen is lost through charge
exchange in the heliosheath regions.  The detailed spectra
characterized by the ENAs provide time-tagged samples of the energy
distributions of the underlying ion distributions, and provide a
wealth of information about the outer heliosphere regions, and beyond.


\end{abstract}

\section{Introduction} \label{intro}

The newest frontier in space exploration is close to home, where the
outflowing magnetized solar wind  plasma meets and mixes with the
interstellar cloud at the heliosphere's boundaries.  The evolving
heliosphere is formed by the relative ram (dynamic) pressures of the
solar wind and partially ionized low density, $\sim 0.3$ \cc,
interstellar cloud flowing around (interstellar ions and magnetic
field) and through (interstellar neutrals) the heliosphere at 26.3
\kms.  The interstellar neutrals dominate the mass-density in the
heliosphere beyond 10--15 AU, and are able to penetrate deep into the
inner heliosphere.  Gravitational focusing of heavy elements in the
flow forms the helium focusing cone that the Earth traverses early
every December.  The Interstellar Boundary Explorer (IBEX) mission
\citep{McComasetal:2009ssr} 
has recently mapped the energetic neutral atoms (ENAs) that are formed
by charge-exchange between heliosphere plasmas and interstellar
neutrals in the heliosheath regions.  These maps are now reshaping our
understanding of the heliospheric interaction with the interstellar
medium.  IBEX has also made the first $in~situ$ detection of the flow of
interstellar oxygen through the heliosphere \citep{Moebius:2009sci}.

The presence of interstellar neutrals in the heliosphere was
established 40 years ago.  OGO 5 mapped the \HI\ Lyman alpha sky
background and showed a diffuse component attributed to interstellar
neutral hydrogen in the inner heliosphere
\citep{ThomasKrassa:1971,BertauxBlamont:1971}. A $Copernicus$ spectrum
of this weak Ly$-\alpha$ emission firmly established it as interstellar
\citep{AdamsFrisch:1977}. Measurements of the fluorescence of solar 584 A
emission from interstellar \HeI\ in the heliosphere revealed the
helium focusing cone \citep{WellerMeier:1974}.  The discovery of
helium pickup ions \citep{Moebiusetal:1985} showed that the
ionization of interstellar neutrals inside of the heliosphere
produces energetic ions that
trace the neutrality of the interstellar cloud around the Sun.
Inside of the heliosphere, interstellar neutrals are ionized through 
charge-exchange with the solar wind, photoionization,
and for those neutrals surviving to 1 AU electron-impacts
\citep{Rucinskietal:1996}.  Pickup ions are a crucial diagnostic
of the interaction between the heliosphere and the interstellar medium.

ENAs formed from charge-exchange between interstellar neutrals and 
solar wind ions were recognized as an important remote
diagnostic of the distant heliosphere boundary regions 
\citep{Gruntman:1993,HsiehGruntman:1993}.
ENAs with energies 55--80 keV, originating in the heliosheath,
were discovered by 
CELIAS/HSTOF on SOHO \citep{Hilchenbachetal:1998ena}.  
IBEX maps of ENAs produced in the heliosphere boundaries,
and the unexpected discovery of the IBEX
Ribbon, requires a new paradigm for understanding the interaction
between the heliosphere and interstellar medium
\citep{McComas:2009sci,Funsten:2009sci,Fuselier:2009sci,Schwadron:2009sci}.

Interstellar neutral helium is the best marker for the upwind direction of the ISM
flowing through the heliosphere, usually denoted the heliosphere nose direction.
\citet{Moebiusetal:2004} combined several data sets to obtain an upwind
direction toward \elon,\elat$ \sim 255^\circ, ~ 5^\circ$ (or in galactic
coordinates, $ \ell \sim 4^\circ, ~ b \sim 15^\circ$).
Photoionization models of the interstellar cloud surrounding the heliosphere,
that are constrained by the observed properties of interstellar material inside
and surrounding the heliosphere, show that the circumheliospheric
interstellar cloud is low density, partially ionized, and warm,
\nHI$\sim 0.20$ \cc, \nel$\sim 0.07$ \cc, \npro$\sim 0.06$ \cc, and $T
\sim 6300$ K \citep[Model 26 in ][]{SlavinFrisch:2008}.  If thermal and
magnetic pressures are similar, the interstellar magnetic field (ISMF)
strength is $\sim 2.7$ \microG.

\begin{figure}[t!]
 \begin{center}
  \includegraphics[width=0.950\textwidth]{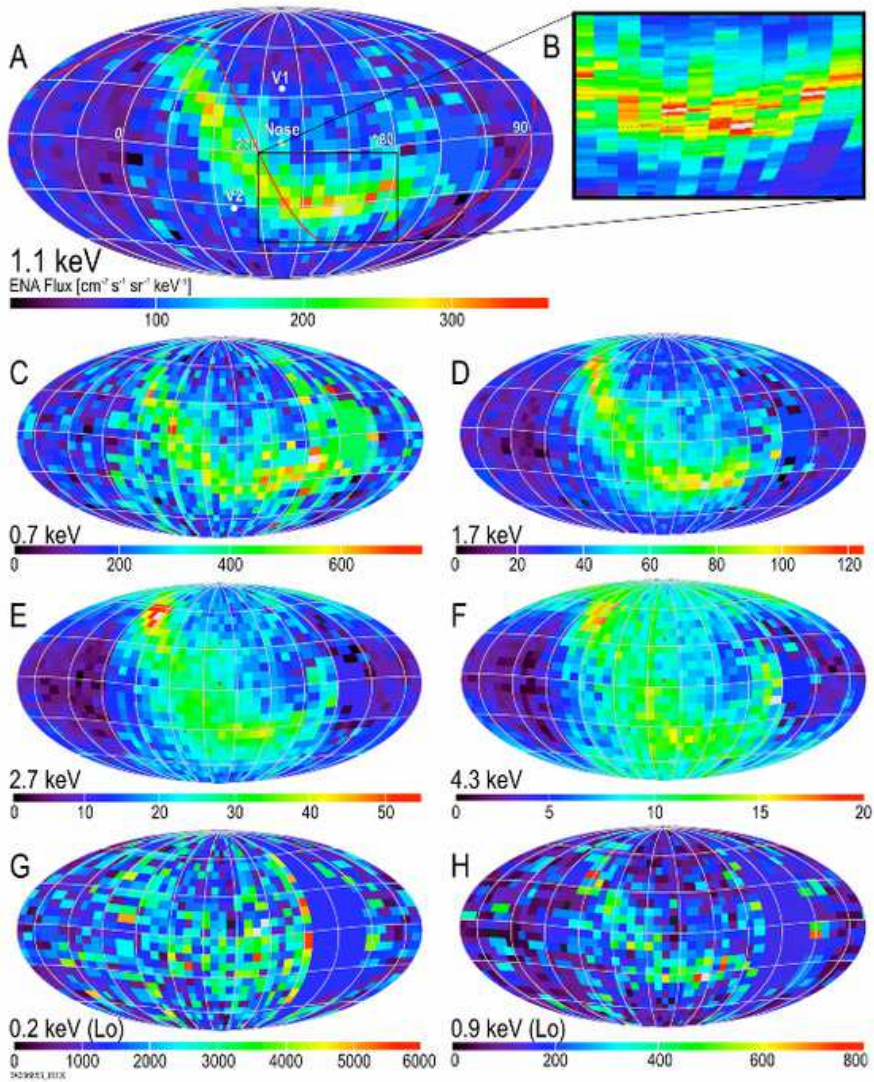}
\end{center}
\caption{ IBEX measurements of the first all-sky maps of ENA fluxes
over central energies of 0.2 keV (Lo) through 4.3 keV (Hi).  The maps are
Molleweide projections in ecliptic coordinates (J2000), and are
centered near the longitude of the heliosphere nose.  Also shown are
the galactic plane (red line in upper figure) and the current
locations of the Voyager 1 (at \elon,\elat=$255^\circ, ~ 34^\circ$)
and Voyager 2 ($289^\circ, -29^\circ$) satellites.  The prominent arc
of bright emission is the IBEX Ribbon.  The Ribbon is not ordered by
either ecliptic or galactic coordinates, but instead appears to form
where the sightlines are perpendicular to the direction of the
interstellar magnetic field draping over the heliosphere.  The
magnified region shows fine scale structure in the Ribbon, which has
been identified by smoothing each $0.5^\circ$ angle in spin phase
along a measurement swath by the amount needed to reach 10\% counting
statistics.  The ENA energies are plotted in the spacecraft
frame. Pixels in the small area near $\lambda \sim 120^\circ -
150^\circ$ are contaminated by the magnetosphere, and are filled in by
extrapolation from adjacent regions. \citep[The figure is from ][]{McComasetal:2010var}.}
\label{fig:1}
\end{figure}


\section{Measurements of Energetic Neutral Atoms from the Heliosphere Boundaries } \label{sec:results}

IBEX was launched October 19, 2008 into low-Earth orbit and
subsequently raised itself into a highly ecliptical orbit, $\sim 2-50
~ R_\mathrm{Earth}$.  The IBEX spacecraft is Sun-pointing, and spins
four times a minute, so that the two oppositely mounted detectors,
IBEX-Hi \citep{Funstenetal:2009ssr} and IBEX-Lo
\citep{Fuselieretal:2009ssr} measure ENAs over great circle swaths of
the sky. The spacecraft is repointed towards the Sun once a week,
yielding an effective angular resolution of $\sim 7^\circ$ in longitude.
IBEX completes two all-sky
ENA maps per year. The first set of skymaps were collected December
2008 through June 2009 \citep{McComas:2009sci} and the second set
from June through December 2009 \citep{McComasetal:2010var}, during
periods when the orbit apogee was primarily outside of the
magnetosphere and in the magnetosheath and solar wind.  IBEX-Lo has
eight energy channels that measure ENAs in the interval $\sim 0.01 - 2
$ keV.  IBEX-Lo also has the capability to directly measure the impact
of neutral atoms with relatively high densities in the surrounding
interstellar cloud, such as H, He, O, and possibly Ne, during certain
portions of the yearly orbit \citep{Moebius:2009sci}.  IBEX-Hi has
six energy channels covering $\sim 0.3 - 6 $ keV.

\subsection{Characteristics of the ENA Sky}

The IBEX all-sky maps of ENA fluxes for IBEX-Hi and
IBEX-Lo energy bands are shown in Figure \ref{fig:1}, in units of \units.  These maps are
centered near the heliosphere nose, located at ecliptic coordinates $\lambda
\sim 255^\circ,~\beta \sim 5^\circ$.  The distribution of ENA
emission on the sky is not ordered by either ecliptic or galactic
coordinates.  The most obvious feature in the ENA sky is the
newly-discovered bright 'Ribbon' of ENA emission.  The Ribbon forms a
nearly complete arc on the sky, centered $\sim 46^\circ$ from the
heliosphere nose toward $\lambda,~\beta \sim 221^\circ,~39^\circ$.
None of the pre-launch ENA models predicted the Ribbon.

\begin{figure}[b!]
\vspace*{-0.2in}
\begin{center}
\includegraphics[width=0.8\textwidth]{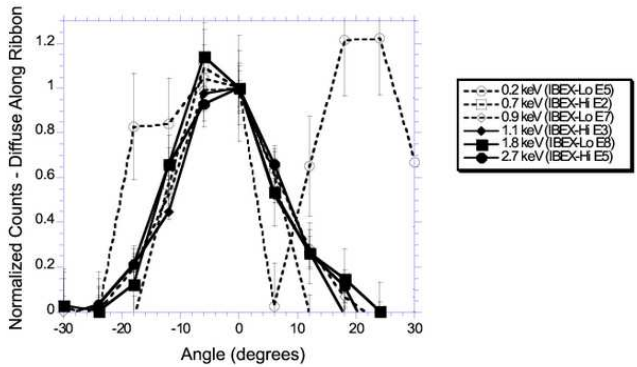}
\end{center}
\vspace*{-0.3in}
\caption{Energy fluxes shown for a cut across the IBEX Ribbon
in directions that are approximately perpendicular to the arc of the Ribbon.
The Ribbon fluxes are normalized by a baseline distributed emission that is
subtracted at each energy, where the baseline flux is a linear polynomial 
that is required to fit the total ENA fluxes well outside 
of the Ribbon arc.  After normalizing the Ribbon fluxes by the relatively
smooth distributed emission, the full-width-half-max of the Ribbon is $\sim 20^\circ$
at each of the IBEX energy bands where the Ribbon is apparent (0.1 keV -- 6 keV),
with the exception of the 1.1 keV channel where fluxes are enhanced at the core
solar wind energy.  The Ribbon fluxes are 2--3 times the fluxes of the distributed
emissions. \citep[This figure is from][]{Fuselier:2009sci}.}
\label{fig:2}
\end{figure}

The IBEX Ribbon is observed between 0.2 and 6 keV, with an average
width that is independent of energy and narrow \citep[$\sim 20^\circ$,
][]{Fuselier:2009sci}.  The Ribbon width is determined by averaging
ENA fluxes over $60^\circ$ segments across the Ribbon, and normalizing the co-added segments by
fitting and subtracting the underlying distributed emission
(Figure \ref{fig:2}).  The excess emissions in both figures
for angles to the Ribbon-center line of $\sim 12-30^\circ$ are counts
from directly detected interstellar neutrals (Section \ref{sec:isn}).
The Ribbon is therefore characterized as a coherent long
and narrow arc in the sky, spanning $\ge 300^\circ$, with a similar spatial
morphology over the IBEX energy range.

The Ribbon contains fine structure, including small regions of bright
ENA emission, or 'hot-spots', that are several pixels wide and of
different lengths \citep{McComas:2009sci}.  The most prominent
hot-spot, referred to as the ``knot'' in the ribbon, is located near
$\lambda \sim 350^\circ ,~\beta \sim 60^\circ $.
The second all-sky map has shown that this knot diminished and spread
out along the Ribbon over the six months between the first and second
set of skymaps \citep{McComasetal:2010var}.

The ages of the ENAs forming a skymap vary because of the energy-dependent
time-lag between the creation of the solar wind (SW) ion that
nucleates the ENA, and the ENA detection by IBEX.
Also, a single skymap is collected over six months.
The round trip ENA travel time to the termination shock, 
nominally at $\sim 100$ AU, is 4.3--1.3 years for IBEX-Hi ESA2 and ESA6 
(when the entire energy channel widths are included).  ENAs originating near
the heliopause ($\sim 160$ AU in the upwind direction), show even larger
possible time-lags of 6.8 -- 2.0 year for round trip travel times.

These travel time-lags suggest that the heliosheath ENAs contributing to the 
first skymaps were produced partly by solar wind emitted during the
extreme solar minimum conditions of the declining phase of the Solar Cycle 23.
The solar wind varies between solar minimum with typical velocities $\sim 400$ \kms,
to solar maximum with velocities $\sim 500$ \kms\ or more.
Therefore tracking the observed ENAs back to the parent ions
will require folding the solar-cycle dependence of the heliosphere
into detailed models for ENA production as a function of energy and location.

\begin{figure}[ht]
\begin{flushleft}
   \includegraphics[width=0.48\textwidth]{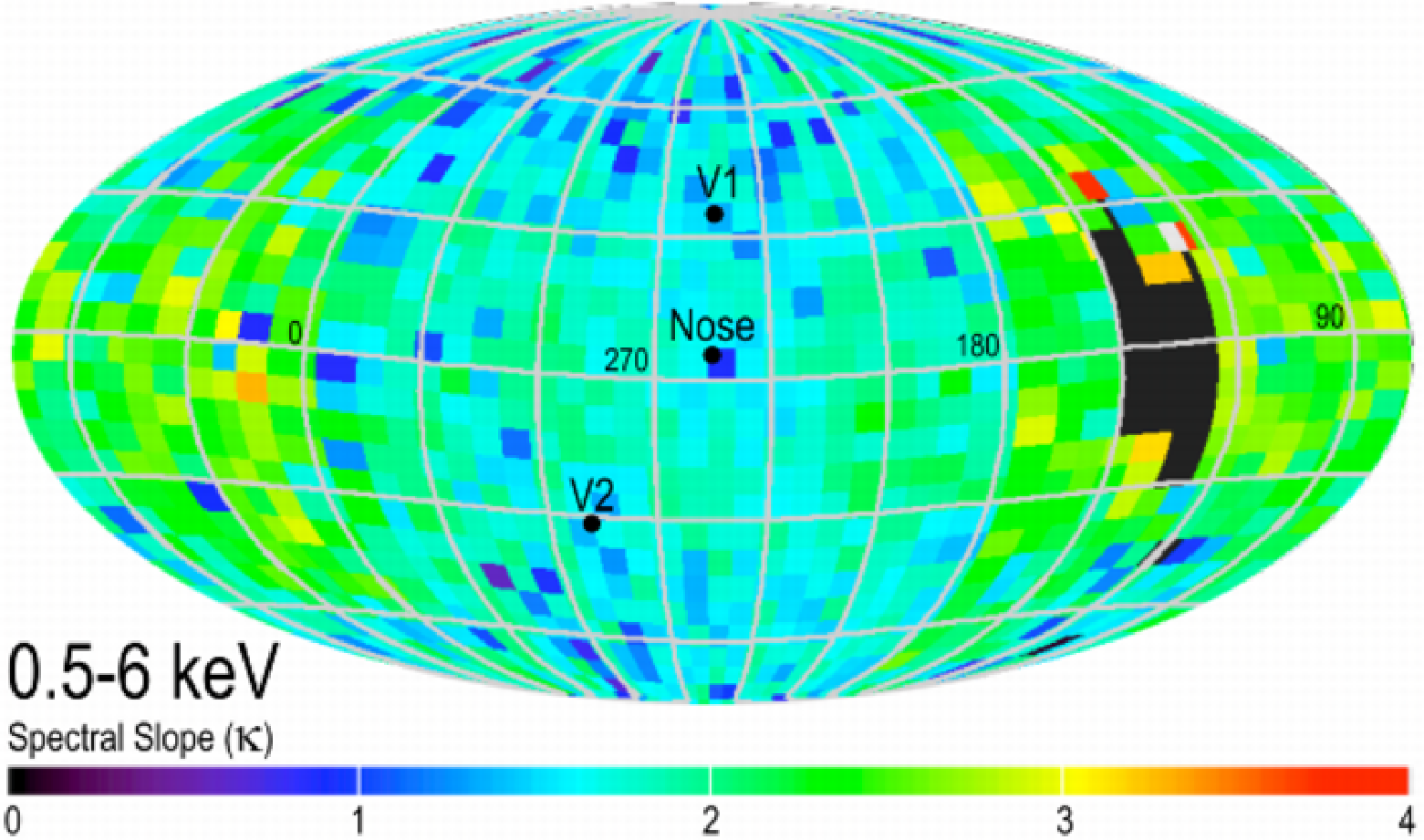}
\end{flushleft}
\hspace*{0.2in}
\vspace*{-1.6in}
\begin{flushright}
   \includegraphics[width=0.48\textwidth]{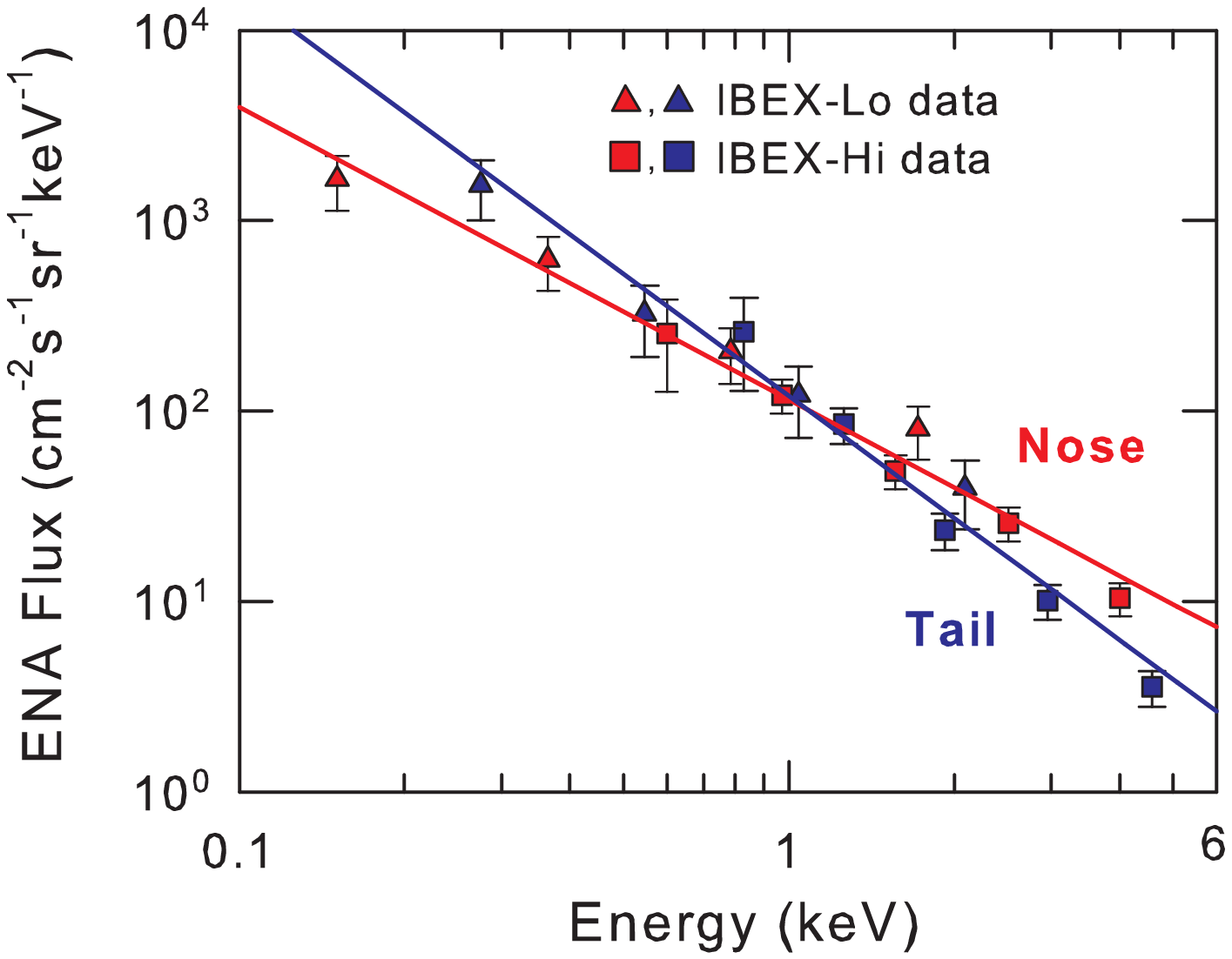}
\end{flushright}
\caption{
Left:  Skymap of the power law slope
of the ENA spectrum, $\kappa$, over the
sky for ENA energies 0.5--6 keV using IBEX-Hi data \citep[from][]{McComas:2009sci}.
The region contaminated by the magnetosphere is left black here.  Right:  The power
law slope of the ENA spectrum in the direction of the heliosphere nose region
(red) and tail region (blue).  Note the softer spectrum towards the tail
\citep[from ][]{Funsten:2009sci}.
}
\label{fig:3}
\end{figure}

\subsection{Energy Distributions of IBEX ENAs}

Fits to nine energy channels 0.7--6 keV show that the ENA spectra
generally follow the power laws that characterize the energy
distribution of the underlying non-thermal heliosheath plasmas
\citep{McComas:2009sci,Funsten:2009sci}.  The spectral index of ENA
emission can be described by a kappa distribution, consisting of a
thermal core and power-law with exponent $\kappa$ at high energies
\citep{Vasyliunas:1968}.  ENA spectra vary with ecliptic longitude
and latitude for energies $> 0.5$ keV (Figure \ref{fig:3}). Outside of
the Ribbon, the globally distributed flux in the north and south polar
regions (\elat$>|54^\circ$) have more complicated spectral shapes that
can not be fit with a single kappa distribution.
For the low latitude spectra, including both the distributed and
Ribbon emission, a power law of $\kappa = 1.5$ is seen towards the
nose regions, with a softer spectra $\kappa = 2.1$ in the direction of
the elongated tail.  
The harder ENA spectra near the polar regions may
result from increased solar wind velocities at high
latitudes.  The softer tail ENA spectrum may show the slowdown of
the solar wind in the tail because of energy loss by charge-exchange
with interstellar neutrals.
For a constant ENA production rate
as a function of energy and distance, the differential cross-sections
would lead to a hardening of the ENA spectra with the distance of the
formation region.  
The charge-exchange cross sections decrease with
increasing energy \citep{LindsayStebbings:2005}, for example varying by a factor
of $\sim 2$ between 0.5 keV (the FWHM lower energy of IBEX-Hi ESA2) and 6.0 keV
(the FWHM upper energy of IBEX-Hi ESA6).  

Surprisingly, the Ribbon does not create an obvious feature in the ENA
energy distribution, even though the fluxes are several times larger
than for adjacent regions.  The spectra of the Ribbon is characterized
by several $\kappa$ values. There is a slight tendency for $\kappa$ to
be slightly softer, $\sim 0.3 $, in the low-latitude Ribbon emission
than in the low-latitude distributed emission \citep[Figure 2A and 2B in][]
{Funsten:2009sci}.  The Ribbon spectra at high-latitude is
similar to the spectra of the distributed emission, with the exception
of the bright emission knot.  At 1.7 and 2.7 keV the knot flux is
highly variable over small spatial scales and shows enhanced fluxes
compared to the distributed emission.  This emission knot is located
toward a region where slow and fast solar winds may interact,
suggesting possible unusual conditions for energizing the pickup ions
for such a termination shock location \citep{McComas:2009sci}.  The
second IBEX skymaps show that the knot emission is time-variable
\citep{McComasetal:2010var}.

\begin{figure}[ht]
\begin{center}
\includegraphics[width=0.455\textwidth]{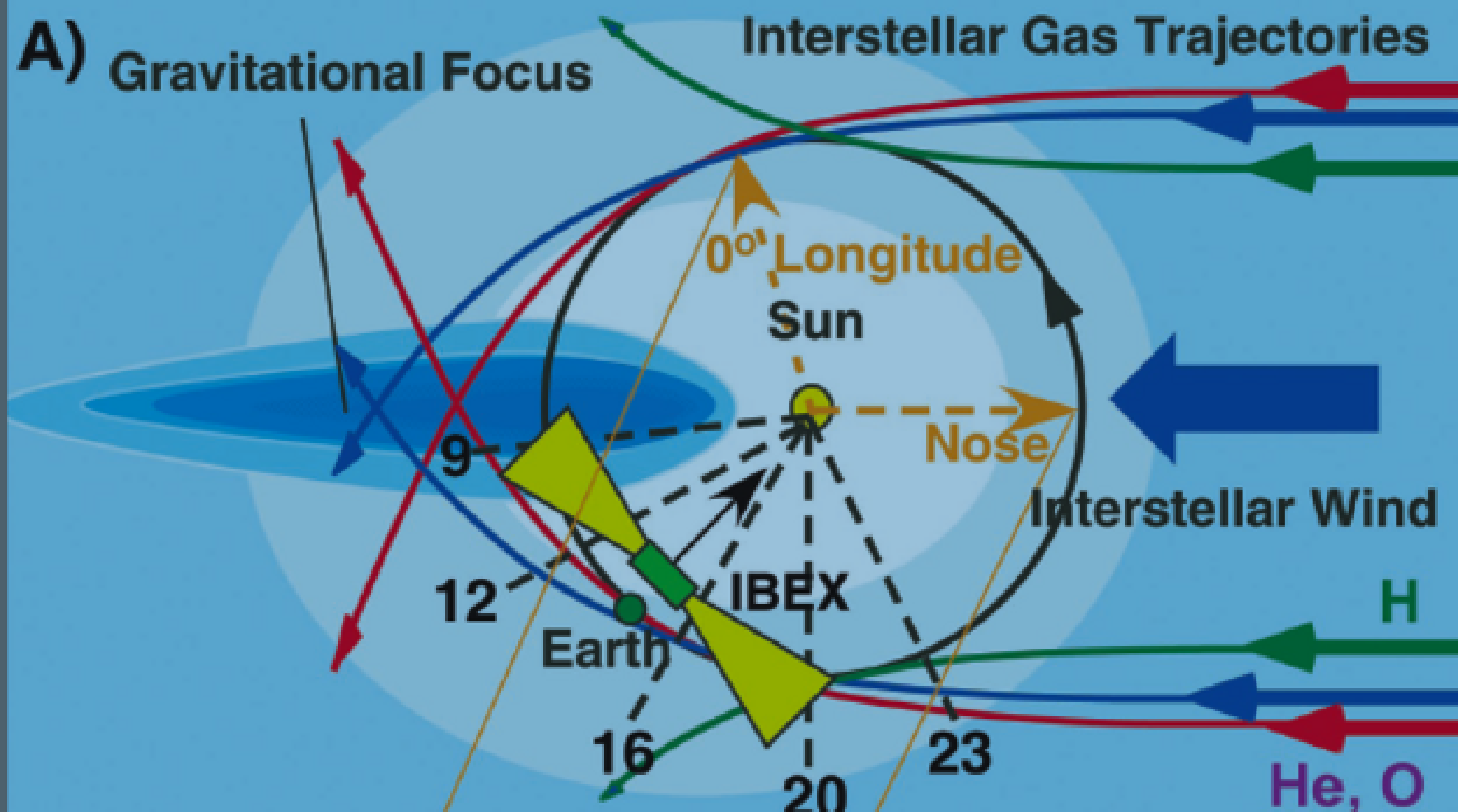}
\includegraphics[angle=90,width=0.455\textwidth]{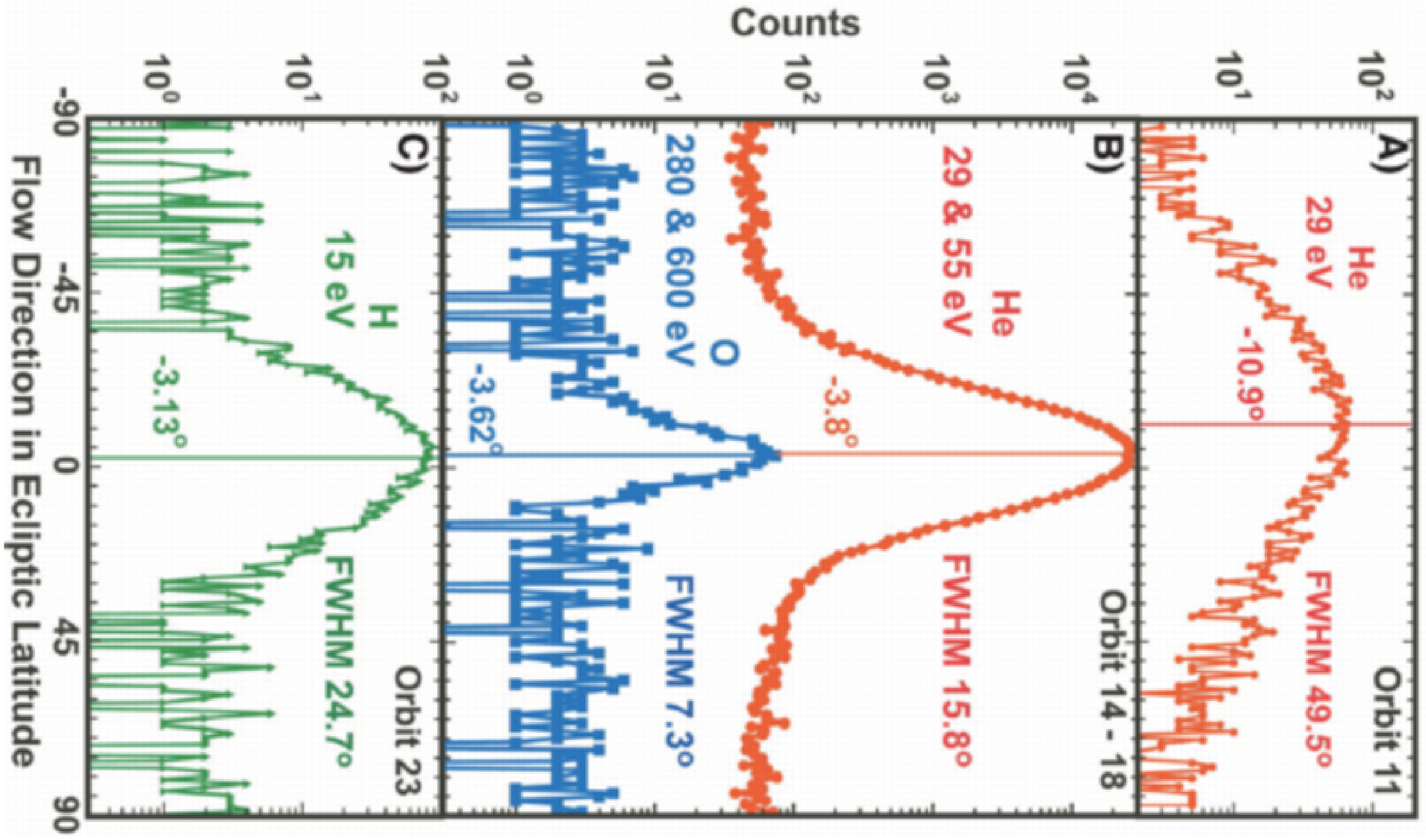}
\end{center}
\caption{Left: Cartoon showing the flow of interstellar neutrals
through the inner heliosphere as deflected by gravity.  The He
focusing cone is shown in blue. Gravitational deflections are stronger
for lighter atoms (He, red) than heavier atoms (O, blue), so that they
are detected during different parts of the orbit.  Hydrogen atoms are
deflected by radiation pressure (green).  The Earth's positions during
IBEX orbits 9--23 is shown, along with the IBEX FOV for orbit 14.
Right:  Direct measurements by IBEX-Lo of the flow of interstellar H (C),
He (A and B), and O (B) are shown as a function of  the ecliptic latitude.  The central and FWHM of each distribution
were found from Gaussian fits to the incoming flow, convoluted with
the IBEX-Lo angular response function. Counts are integrated over
$1^\circ$ bins.  \citep[These figures are from][]{Moebius:2009sci}.  }
\label{fig:4}
\end{figure}

\section{Measuring the Flow of Interstellar Neutrals through the Inner
Heliosphere} \label{sec:isn}

IBEX-Lo has the capability to expand the sample of cosmic material
that has been detected in situ by spacecraft, and provide volume
densities for key interstellar elements.  Several elements in the
circumheliospheric interstellar material (ISM) have a significant
fraction of interstellar neutrals (ISNs), such as H ($\sim 78$\%), He
($\sim 61$\%), O ($\sim 81$\%) and Ne \citep[$\sim 20$\%, Model 26 in
][]{SlavinFrisch:2008}.  Hydrogen and oxygen have similar first
ionization potentials and are tightly coupled in the local ISM.  The
volume densities of ISNs inside of the heliosphere are related to the
ionization state of the circumheliospheric ISM, once filtration in the
heliosheath regions is included.  Volume densities are
difficult to obtain astronomically in general, and almost impossible
to obtain for low density clouds such as the one that surrounds the Sun, making the
IBEX-Lo measurements a valuable diagnostic of the ionization level and
physical properties of the surrounding ISM.

The ISNs enter the heliosphere at interstellar velocities ( $\sim
26.3$ \kms) and follow Keplerian orbits modified by gravity and
radiation pressure into the inner heliosphere.  IBEX-Lo includes the
capability to directly image the flow of several interstellar neutrals
during parts of the year where gravitational deflection and orbital
motions boost neutral velocities above the IBEX-Lo detection
threshold of 0.01 keV.  These directly detected ISNs create peaks in
count rates in the IBEX-Lo sky-maps, at element-specific orbits where
the required energy boost is sufficient \citep[e.g. Figure 1 in ][]{Moebius:2009sci}.
Interstellar H and He create
pronounced features in the IBEX-Lo maps at 0.015 keV and 0.11 keV, and the weaker O
enhancement is visible at 0.6 keV.
All three neutrals arrive from a direction
that appears consistent with the inflowing \HeI\ direction measured by Ulysses
\citep{Moebiusetal:2004}, slightly above the ecliptic plane.  The IBEX scan
strategy gives the distribution of interstellar \HI, \HeI, and \OI\ as
a function of ecliptic latitude (Figure \ref{fig:4}).  The ISN
particle trajectories are reconstructed using a hot-model of the ISM
flow, incident velocities of 26.3 \kms, and ISN gas temperatures of
6,300 K.  The oxygen latitudinal distribution shows an asymmetry, with
a weak enhancement down to \elat$\sim -30^\circ$, that suggests that a
secondary component of the interstellar O flow forms from the ionized
O that is deflected around the heliopause together with the deflected
hydrogen.  A secondary He component is possible \citep[][private communication]{Moebius:2009sci}.

\begin{figure}[th]
 \begin{center}
 \includegraphics[width=0.635\textwidth]{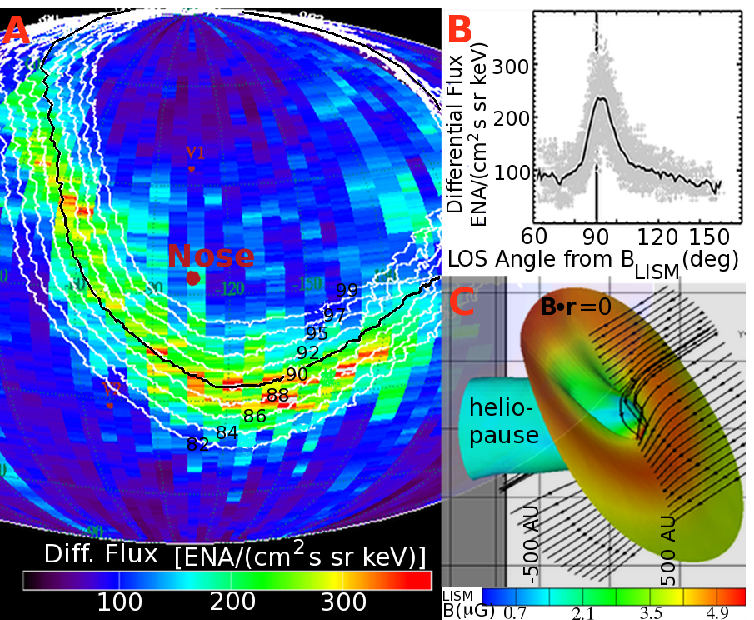}
\end{center}
\caption{(A) The contours of $B \cdot R $ predicted by MHD models
of the heliosphere \citep{Schwadron:2009sci,Pogorelovetal:2009L}) are
superimposed on the 1 keV ENA map.  The black contour shows $B \cdot R
\sim 0$, where the ISMF is perpendicular to the sightline $R$. The
contour labels give the angle between the sightline and $B_\mathrm{LISM}$. 
(B) Plot of the differential ENA flux at 1 keV versus the $B \cdot R$
angle, with the Ribbon emission peak at an angle of 90\deeg\ clearly
visible.  (C) The global configuration of the ISMF field lines draped
over the heliopause, and the surface where $B \cdot R \sim 0$ (red),
and is the 3D realization of the black contour in (A) \citep[figure
from][]{Schwadron:2009sci}.}
\label{fig:5}
\end{figure}

\section{Whence the ENA Ribbon?}\label{sec:interp}

IBEX's discovery of a Ribbon of ENA emission was entirely unexpected,
but at the same time the fundamental directions traced by the Ribbon are
consistent with pre-launch MHD heliosphere models that reproduce known
asymmetries \citep{Schwadron:2009sci,Pogorelovetal:2009L}.  The
asymmetries include the 10 AU difference between termination shock
encounters by Voyager 1 (34\deeg\ above the ecliptic plane, Figure 1)
and Voyager 2 (29\deeg\ below the plane), the 5\deeg\ offset between
the inflow directions of interstellar \HI\ and \HeI\ into the
heliosphere \citep[after correction to J2000 coordinates,][]{Lallementetal:2005,Frisch:2010s1}, and the distribution along
the galactic plane of 3 kHz emissions detected by the Voyager
satellites during the early 1990's \citep{KurthGurnett:2003}.  The
observed asymmetries impose the restriction that the direction of the
interstellar magnetic field is near the plane formed by
inflowing \HI\ and \HeI\ velocity vectors, the 'hydrogen deflection
plane', and that it makes some angle, e.g. $\sim 45^\circ$, with the
gaseous flow velocity
\citep{Pogorelovetal:2009L,OpherRichardson:2009}.  Figure \ref{fig:5} (A)
compares the 1 keV Ribbon fluxes with $B \cdot R \sim 0 $
\citep[from the MHD heliosphere model of ][]{Pogorelovetal:2009L}
which includes ion-neutral coupling with a kinetic description
of the neutrals and Lorentzian description of the ions.
A plot of the fluxes versus $B \cdot R $ is shown in (B), while
(C) shows a 3D figure of the ISMF lines draping over the heliosphere,
together with the $B \cdot R = 0$ surface.
The ISMF-driven asymmetry is offset by the symmetrizing effect of charge-exchange
with the partially ionized low density cloud around the Sun (\nHI$\sim
0.2$ \cc, \nHII$\sim 0.07$ \cc, T$\sim 6,300$ K).  Remarkably, the
centroid ridge of the IBEX Ribbon was found to track directions where
the Pogorelov et al.  heliosphere model indicates that the ISMF
draping over the heliosphere is perpendicular to the sightline (Figure \ref{fig:5}, A).
The central direction of the Ribbon arc, towards
galactic coordinates L,B$=33^\circ,55^\circ$ \citep[or ecliptic coordinates
$\lambda,\beta = 221^\circ,39^\circ$, ][]{Funsten:2009sci}
is within the $\pm 35^\circ$ uncertainties of the best fitting
local ISM magnetic field direction of L,B$=38^\circ,23^\circ$, found
from starlight polarizations caused by
magnetically aligned interstellar dust grains \citep{Frischetal:2010pol}.
The discovery of the IBEX Ribbon requires rethinking our paradigm of how
the solar wind interacts with interstellar gas in the surrounding
cloud, a process that fundamentally determines the characteristics of
the heliosphere.

\begin{figure}[ht]
 \begin{center}
  \includegraphics[width=0.935\textwidth]{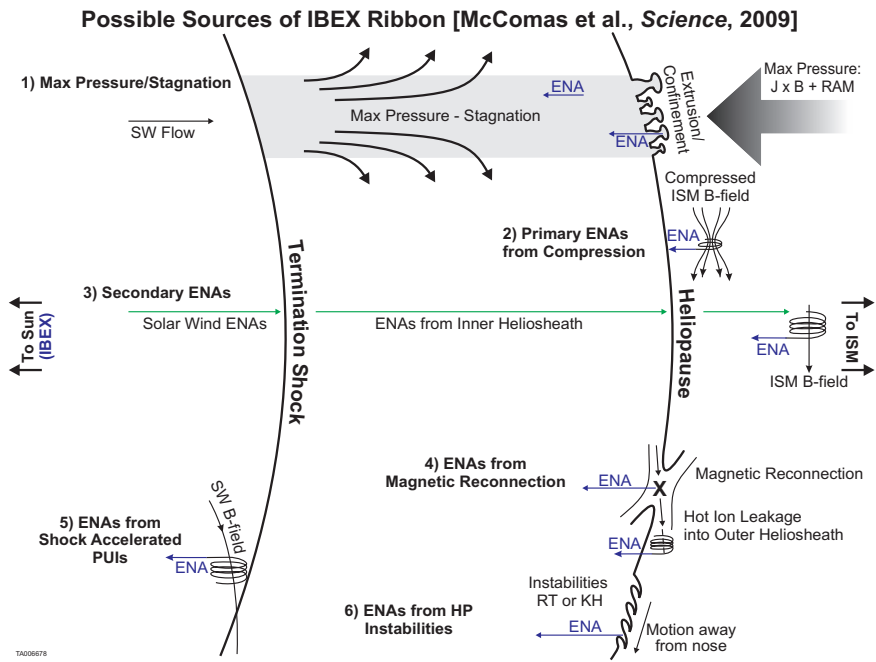}
\end{center}
\caption{ Different possible sources of the enhanced ENA emission that
forms the IBEX Ribbon are summarized in this cartoon.  Possibility (1)
generates ENA production in the regions of high pressure and plasma
stagnation in the inner heliosheath, and possibly associated with
heliopause distortions that create outward extrusions of plasma.
Process (2) corresponds to ENAs produced preferentially with 90\deeg\
pitch angles in the interstellar magnetic field that is compressed
against the heliopause.  The Lorentz force then provides the mechanism
for diverting the particle momentum back to the inner heliosphere to
be collected by IBEX. The third mechanism (3) creates secondary ENAs,
in the inner heliosheath or inside the termination shock, through
ionization and subsequent remission of outward traveling ENAs created
from the solar wind and pickup ions.  (4) The ENAs created from
magnetic reconnection across the heliopause, which mixes inner
heliosheath ions with neutrals, will show the direct imprint of the
ISMF.  (5) ENAs may be produced from shock accelerated ions close to
the termination shock in the upstream and downstream regions.
Possibility (6) produces ENAs in transient regions at the heliopause
resulting from Rayleigh-Taylor or Kelvin-Helmholz instabilities. This
figure is from McComas et al. 2010.}
\label{fig:6}
\end{figure}

Several possible mechanisms have been suggested for the formation of
the Ribbon \citep{McComas:2009sci,McComasetal:2010var}.
Figure \ref{fig:6} \citep{McComasetal:2010var} illustrates possible scenarios 
suggested initially by the IBEX team.

(1) The Ribbon forms in inner heliosheath regions of maximum pressure,
so that plasma flows away from the Ribbon (consistent with Voyager
data) and the plasma stagnates.  The pattern of ISM pressure (dynamic
and magnetic) would imprint on the subsonic inner heliosheath, and
those heliosheath regions with higher densities would emit more ENAs.
Outwards extrusions, through the heliopause, of pockets of higher
density gas might generate similar ENA spectra as the inner
heliosheath.  \citet{Schwadron:2009sci} and \citet{Fuselier:2009sci} estimate a
heliosheath thickness of $L \sim 50$ AU if the observed width and
thickness of the Ribbon are comparable.  
\citet{Hsiehetal:2010ena} calculate ENA production from 0.04--4 MeV
ions in the heliosheath measured by Voyager 1 and Voyager 2,
and obtained a thicker Ribbon in the
southern ecliptic hemisphere, $L = 28 \pm 8$ AU, than in the
northern hemisphere, $L = 21 \pm 6$ AU, from comparisons with
ENA data up to 0.1 MeV from SOHO/HSTOF.  The larger
heliosheath thickness inferred from the 44 keV $Cassini/INCA$
\citep{Krimigis:2009sci} data may point to a more distant source region 
for these high energy ENAs.  If the ring of higher interstellar pressure associated with the Ribbon is
then felt inside of the heliopause, the large scale pressure driving
the Ribbon is an external effect, and so would be expected to result
in a largely stable feature \citep{McComasetal:2010var}.

(2) The Ribbon forms in outer heliosheath regions,possible from secondary ENAs (see (3)),
where the ISMF is compressed against the heliopause.
For this scenario, solar cycle variations should have less impact on the Ribbon
than for (1).  Three characteristics stand out for a 
Ribbon configuration originating beyond the heliopause, based on the MHD
heliosphere model used in \citet[][from Pogorelov et al., 2009]{Schwadron:2009sci}. 
(a) The width of the Ribbon is controlled by the limits set on $|B \cdot
R|$.  (b) The length of the Ribbon arc is set by the limits set on
total magnetic pressure, $\sim {B}^2$.  (c) The latitude of the Ribbon
arc, compared to the magnetic pole defined by the center of the
arc, depends on the distance beyond the heliosphere of the Ribbon
formation, so that a more distant ENA source location will slide the
Ribbon towards the equator of the ISMF traced by the Ribbon arc.

(3) The third possible mechanism forms the Ribbon beyond the
heliopause from secondary ENAs.   
Primary ENAs are produced from neutral charge-exchange with solar wind or inner heliosheath
pickup ions, and propagate outwards past the heliopause
where they convert back to energetic ions through charge exchange.  These ions
convert, via charge-exchange, to the secondary ENAs that propagate 
inwards to IBEX \citep{Heerikhuisen:2010ribbon,Florinskietal:2010,Chalovetal:2010ribbon}.
The mean free path of a 100 (400) \kms\ ENA in the outer
heliosheath is $\sim 250 ~ (400)$ AU,
suggesting that Ribbon ENAs must form over lengths of $\ge 200$ AU.  The
long times (decade) for parent ions to travel outwards and return as
ENAs might be expected to smooth out solar cycle variations and small
scale structures in the ENA fluxes.

(4) ENAs from magnetic reconnection at the heliopause: Magnetic
reconnection is a mechanism that directly mixes the interstellar and
subsonic solar plasmas, and might form transient structures in the
Ribbon in the regions where the interstellar pressure is at a maximum.
Alternating solar polarities would disperse the reconnection across
both hemispheres of the heliosphere, but a continuous Ribbon
structure would need explaining.

(5) If the subsonic solar plasma disperses the increased interstellar
pressure throughout the inner heliosheath, the termination shock
location may also be affected, and therefore, the production or
energization of pickup ions at the termination shock may increase.

(6) The dense interstellar plasma and tenuous time-variable solar wind
plasma are separated by the tangential discontinuity known as the
heliopause.  The heliopause is generally thought to be unstable,
either from shear instabilities or the effective gravity of
charge-exchange \citep[e.g.][]{Dasgupta:2006heliopauseinstability}.
Rayleigh-Taylor instabilities are expected to dominate near the
heliosphere nose, and shear (Kelvin-Helmholz) instabilities near the
heliosphere flanks.  These instabilities directly mix the inner
heliosheath solar wind and pickup ion plasmas with the interstellar
neutrals and plasmas, and thereby could increase the ENA production
rate (e.g. by reducing hydrogen filtration).

There is also at least one entirely new origin suggested
for the Ribbon, whereby it is a viewpoint perspective of ENAs formed
where ISNs in the interface on the circumheliospheric interstellar gas
charge-exchange with the surrounding hot plasma, $\sim 10^6$ K
\citep{GrzedzielskiBzowski:2010ribbon}.  
Despite the strong agreement
of the Ribbon location with the MHD models that reproduce the draping
of the ISMF over the heliosphere, the same field will affect the cloud
around the Sun where it may create an analogous yet-unknown asymmetry.

\section{Closing Comments}

IBEX continues to return new ENA data that monitor the interaction
between the solar wind and interstellar medium at the heliosphere
boundaries.  Two full sky maps have been completed and initially
analyzed, and they show firmly that the fluxes of ENAs formed at the
heliosphere boundaries vary with the solar magnetic activity cycle
\citep{McComasetal:2010var}.  The potential for ENAs to trace the relation
between stellar winds and interstellar gas is enormous, with
implications for the structure of the heliosphere and astrospheres
around exoplanetary systems.  IBEX is also expected to detect the
solar transition between interstellar clouds if it occurs 'soon'
\citep{Frischetal:2010}. As data builds up over time on the population of
interstellar neutrals that are detected by IBEX-Lo, such as hydrogen,
helium and oxygen,
the ability to develop self-consistent models of the
physical properties of the ISM will also improve. IBEX data bridges
the gap between solar system and Milky Way Galaxy by mapping the
energetic neutral atoms that form from charge-exchange between
interstellar neutral hydrogen atoms and the solar wind and pickup
ions, and direct measurements of interstellar
neutrals \citep{Frischetal:2010}.


\begin{acknowledgements}
We thank the IBEX team members 
for the exciting discoveries of the IBEX mission, and
the International Space Sciences Institute in Bern, Switzerland, for
hosting the workshop 'Galactic Cosmic Rays in the Heliosphere'
where this review was presented.  
This work was funded through the IBEX mission, as a part of NASA's
Explorer Program.
\end{acknowledgements}


\begin{thebibliography}{36}
\ifx \bisbn   \undefined \def \bisbn  #1{ISBN #1}\fi
\ifx \binits  \undefined \def \binits#1{#1} \fi
\ifx \bauthor  \undefined \def \bauthor#1{#1} \fi
\ifx \bjtitle  \undefined \def \bjtitle#1{\textrm{#1}}\fi
\ifx \batitle  \undefined \def \batitle#1{#1} \fi
\ifx \bctitle  \undefined \def \bctitle#1{#1} \fi
\ifx \bvolume  \undefined \def \bvolume#1{\textbf{#1}}\fi
\ifx \byear  \undefined \def \byear#1{#1} \fi
\ifx \bissue  \undefined \def \bissue#1{#1} \fi
\ifx \bfpage  \undefined \def \bfpage#1{#1} \fi
\ifx \blpage  \undefined \def \blpage #1{#1} \fi
\ifx \burl  \undefined \def \burl#1{#1} \fi
\ifx \doiurl  \undefined \def \doiurl#1{#1} \fi
\ifx \betal  \undefined \def \betal{et al.} \fi
\ifx \binstitute  \undefined \def \binstitute#1{#1} \fi
\ifx \beditor  \undefined \def \beditor#1{#1} \fi
\ifx \bpublisher  \undefined \def \bpublisher#1{#1} \fi
\ifx \bbtitle  \undefined \def \bbtitle#1{\textit{#1}} \fi
\ifx \bedition  \undefined \def \bedition#1{#1} \fi
\ifx \bseriesno  \undefined \def \bseriesno#1{#1} \fi
\ifx \blocation  \undefined \def \blocation#1{#1} \fi
\ifx \bsertitle  \undefined \def \bsertitle#1{#1} \fi
\ifx \bsnm \undefined \def \bsnm#1{#1} \fi
\ifx \bsuffix \undefined \def \bsuffix#1{#1} \fi
\ifx \bparticle \undefined \def \bparticle#1{#1} \fi
\ifx \barticle \undefined \def \barticle#1{#1} \fi
\ifx \botherref \undefined \def \botherref #1{#1} \fi
\ifx \url \undefined \def \url#1{#1} \fi
\ifx \bchapter \undefined \def \bchapter#1{#1} \fi
\ifx \bbook \undefined \def \bbook#1{#1} \fi
\ifx \bcomment \undefined \def \bcomment#1{#1} \fi
\ifx \oauthor \undefined \def \oauthor#1{#1} \fi
\ifx \citeauthoryear \undefined \def \citeauthoryear#1{#1} \fi
\ifx \texttildelow  \undefined \def \texttildelow{\symbol{126}} \fi
\def \endbibitem {}

\bibitem[\protect\citeauthoryear{{Adams} and {Frisch}}{1977}]{AdamsFrisch:1977}
\begin{barticle}
\bauthor{\binits{T.F.} \bsnm{{Adams}}}, \bauthor{\binits{P.C.}
  \bsnm{{Frisch}}},
\batitle{{High-resolution observations of the \protect{{L}yman} alpha sky
  background}}.
\bjtitle{\apj}
\bvolume{212},
\bfpage{300}--\blpage{308}
(\byear{1977})
\end{barticle}
\endbibitem

\bibitem[\protect\citeauthoryear{{Bertaux} and
  {Blamont}}{1971}]{BertauxBlamont:1971}
\begin{barticle}
\bauthor{\binits{J.L.} \bsnm{{Bertaux}}}, \bauthor{\binits{J.E.}
  \bsnm{{Blamont}}},
\batitle{{Evidence for a Source of an Extraterrestrial Hydrogen Lyman-alpha
  Emission}}.
\bjtitle{\aap}
\bvolume{11},
\bfpage{200}
(\byear{1971})
\end{barticle}
\endbibitem

\bibitem[\protect\citeauthoryear{{Chalov} et~al.}{2010}]{Chalovetal:2010ribbon}
\begin{barticle}
\bauthor{\binits{S.V.} \bsnm{{Chalov}}}, \bauthor{\binits{D.B.}
  \bsnm{{Alexashov}}}, \bauthor{\binits{D.} \bsnm{{McComas}}},
  \bauthor{\binits{V.V.} \bsnm{{Izmodenov}}}, \bauthor{\binits{Y.G.}
  \bsnm{{Malama}}}, \bauthor{\binits{N.} \bsnm{{Schwadron}}},
\batitle{{Scatter-free Pickup Ions beyond the Heliopause as a Model for the
  Interstellar Boundary Explorer Ribbon}}.
\bjtitle{\apjl}
\bvolume{716},
\bfpage{99}--\blpage{102}
(\byear{2010})
\end{barticle}
\endbibitem

\bibitem[\protect\citeauthoryear{{Dasgupta}
  et~al.}{2006}]{Dasgupta:2006heliopauseinstability}
\begin{botherref}
\oauthor{\binits{B.} \bsnm{{Dasgupta}}}, \oauthor{\binits{V.}
  \bsnm{{Florinski}}}, \oauthor{\binits{J.} \bsnm{{Heerikhuisen}}},
  \oauthor{\binits{G.P.} \bsnm{{Zank}}},
{MHD Instabilities at the Heliopause},
in \textit{Physics of the Inner Heliosheath},
ed. by {J.~Heerikhuisen, V.~Florinski, G.~P.~Zank, \& N.~V.~Pogorelov }.
American Institute of Physics Conference Series,
vol. 858,
2006,
pp. 51--57
\end{botherref}
\endbibitem

\bibitem[\protect\citeauthoryear{{Florinski} et~al.}{2010}]{Florinskietal:2010}
\begin{barticle}
\bauthor{\binits{V.} \bsnm{{Florinski}}}, \bauthor{\binits{G.P.}
  \bsnm{{Zank}}}, \bauthor{\binits{J.} \bsnm{{Heerikhuisen}}},
  \bauthor{\binits{Q.} \bsnm{{Hu}}}, \bauthor{\binits{I.} \bsnm{{Khazanov}}},
\batitle{{Stability of a Pickup Ion Ring-beam Population in the Outer
  Heliosheath: Implications for the IBEX Ribbon}}.
\bjtitle{\apj}
\bvolume{719},
\bfpage{1097}--\blpage{1103}
(\byear{2010})
\end{barticle}
\endbibitem

\bibitem[\protect\citeauthoryear{{Frisch}}{2010}]{Frisch:2010s1}
\begin{barticle}
\bauthor{\binits{P.C.} \bsnm{{Frisch}}},
\batitle{{The S1 Shell and Interstellar Magnetic Field and Gas Near the
  Heliosphere}}.
\bjtitle{\apj}
\bvolume{714},
\bfpage{1679}--\blpage{1688}
(\byear{2010})
\end{barticle}
\endbibitem

\bibitem[\protect\citeauthoryear{{Frisch} et~al.}{2010a}]{Frischetal:2010}
\begin{barticle}
\bauthor{\binits{P.C.} \bsnm{{Frisch}}}, \bauthor{\binits{J.}
  \bsnm{{Heerikhuisen}}}, \bauthor{\binits{N.V.} \bsnm{{Pogorelov}}},
  \bauthor{\binits{B.} \bsnm{{DeMajistre}}}, \bauthor{\binits{G.B.}
  \bsnm{{Crew}}}, \bauthor{\binits{H.O.} \bsnm{{Funsten}}},
  \bauthor{\binits{P.H.} \bsnm{{Janzen}}}, \bauthor{\binits{D.J.}
  \bsnm{{McComas}}}, \bauthor{\binits{E.} \bsnm{{Moebius}}},
  \bauthor{\binits{H.R.} \bsnm{{Mueller}}}, \bauthor{\binits{D.B.}
  \bsnm{{Reisenfeld}}}, \bauthor{\binits{J.D.} \bsnm{{Schwadron}}
  \bsuffix{N.~A.~{Slavin}}}, \bauthor{\binits{G.P.} \bsnm{{Zank}}},
\batitle{{Can IBEX Identify Variations in the Galactic Environment of the Sun
  using Energetic Neutral Atom (ENAs)?}}
\bjtitle{\apj}
\bvolume{719},
\bfpage{1984}--\blpage{1992}
(\byear{2010}a)
\end{barticle}
\endbibitem

\bibitem[\protect\citeauthoryear{{Frisch} et~al.}{2010b}]{Frischetal:2010pol}
\begin{barticle}
\bauthor{\binits{P.C.} \bsnm{{Frisch}}}, \bauthor{\binits{B.}
  \bsnm{{Andersson}}}, \bauthor{\binits{A.} \bsnm{{Berdyugin}}},
  \bauthor{\binits{H.O.} \bsnm{{Funsten}}}, \bauthor{\binits{M.}
  \bsnm{{Magalhaes}}}, \bauthor{\binits{D.J.} \bsnm{{McComas}}},
  \bauthor{\binits{V.} \bsnm{{Piirola}}}, \bauthor{\binits{N.A.}
  \bsnm{{Schwadron}}}, \bauthor{\binits{J.D.} \bsnm{{Slavin}}},
  \bauthor{\binits{S.J.} \bsnm{{Wiktorowicz}}},
\batitle{{Comparisons of the Interstellar Magnetic Field Directions obtained
  from the IBEX Ribbon and Interstellar Polarizations}}.
\bjtitle{\apj}
\bvolume{724},
\bfpage{1473}--\blpage{1479}
(\byear{2010}b)
\end{barticle}
\endbibitem

\bibitem[\protect\citeauthoryear{{Funsten} et~al.}{2009a}]{Funsten:2009sci}
\begin{barticle}
\bauthor{\binits{H.O.} \bsnm{{Funsten}}}, \bauthor{\binits{F.}
  \bsnm{{Allegrini}}}, \bauthor{\binits{G.B.} \bsnm{{Crew}}},
  \bauthor{\binits{R.} \bsnm{{DeMajistre}}}, \bauthor{\binits{P.C.}
  \bsnm{{Frisch}}}, \bauthor{\binits{S.A.} \bsnm{{Fuselier}}},
  \bauthor{\binits{M.} \bsnm{{Gruntman}}}, \bauthor{\binits{P.}
  \bsnm{{Janzen}}}, \bauthor{\binits{D.J.} \bsnm{{McComas}}},
  \bauthor{\binits{E.} \bsnm{{M{\"o}bius}}}, \bauthor{\binits{B.}
  \bsnm{{Randol}}}, \bauthor{\binits{D.B.} \bsnm{{Reisenfeld}}},
  \bauthor{\binits{E.C.} \bsnm{{Roelof}}}, \bauthor{\binits{N.A.}
  \bsnm{{Schwadron}}},
\batitle{{Structures and Spectral Variations of the Outer Heliosphere in IBEX
  Energetic Neutral Atom Maps}}.
\bjtitle{Science}
\bvolume{326},
\bfpage{964}
(\byear{2009}a)
\end{barticle}
\endbibitem

\bibitem[\protect\citeauthoryear{{Funsten} et~al.}{2009b}]{Funstenetal:2009ssr}
\begin{barticle}
\bauthor{\binits{H.O.} \bsnm{{Funsten}}}, \bauthor{\binits{F.}
  \bsnm{{Allegrini}}}, \bauthor{\binits{P.} \bsnm{{Bochsler}}},
  \bauthor{\binits{G.} \bsnm{{Dunn}}}, \bauthor{\binits{S.} \bsnm{{Ellis}}},
  \bauthor{\binits{D.} \bsnm{{Everett}}}, \bauthor{\binits{M.J.}
  \bsnm{{Fagan}}}, \bauthor{\binits{S.A.} \bsnm{{Fuselier}}},
  \bauthor{\binits{M.} \bsnm{{Granoff}}}, \bauthor{\binits{M.}
  \bsnm{{Gruntman}}}, \bauthor{\binits{A.A.} \bsnm{{Guthrie}}},
  \bauthor{\binits{J.} \bsnm{{Hanley}}}, \bauthor{\binits{R.W.}
  \bsnm{{Harper}}}, \bauthor{\binits{D.} \bsnm{{Heirtzler}}},
  \bauthor{\binits{P.} \bsnm{{Janzen}}}, \bauthor{\binits{K.H.}
  \bsnm{{Kihara}}}, \bauthor{\binits{B.} \bsnm{{King}}}, \bauthor{\binits{H.}
  \bsnm{{Kucharek}}}, \bauthor{\binits{M.P.} \bsnm{{Manzo}}},
  \bauthor{\binits{M.} \bsnm{{Maple}}}, \bauthor{\binits{K.}
  \bsnm{{Mashburn}}}, \bauthor{\binits{D.J.} \bsnm{{McComas}}},
  \bauthor{\binits{E.} \bsnm{{Moebius}}}, \bauthor{\binits{J.} \bsnm{{Nolin}}},
  \bauthor{\binits{D.} \bsnm{{Piazza}}}, \bauthor{\binits{S.} \bsnm{{Pope}}},
  \bauthor{\binits{D.B.} \bsnm{{Reisenfeld}}}, \bauthor{\binits{B.}
  \bsnm{{Rodriguez}}}, \bauthor{\binits{E.C.} \bsnm{{Roelof}}},
  \bauthor{\binits{L.} \bsnm{{Saul}}}, \bauthor{\binits{S.} \bsnm{{Turco}}},
  \bauthor{\binits{P.} \bsnm{{Valek}}}, \bauthor{\binits{S.} \bsnm{{Weidner}}},
  \bauthor{\binits{P.} \bsnm{{Wurz}}}, \bauthor{\binits{S.} \bsnm{{Zaffke}}},
\batitle{{The Interstellar Boundary Explorer High Energy (IBEX-Hi) Neutral Atom
  Imager}}.
\bjtitle{Space Science Reviews}
\bvolume{146},
\bfpage{75}--\blpage{103}
(\byear{2009}b)
\end{barticle}
\endbibitem

\bibitem[\protect\citeauthoryear{{Fuselier}
  et~al.}{2009a}]{Fuselieretal:2009ssr}
\begin{barticle}
\bauthor{\binits{S.A.} \bsnm{{Fuselier}}}, \bauthor{\binits{P.}
  \bsnm{{Bochsler}}}, \bauthor{\binits{D.} \bsnm{{Chornay}}},
  \bauthor{\binits{G.} \bsnm{{Clark}}}, \bauthor{\binits{G.B.} \bsnm{{Crew}}},
  \bauthor{\binits{G.} \bsnm{{Dunn}}}, \bauthor{\binits{S.} \bsnm{{Ellis}}},
  \bauthor{\binits{T.} \bsnm{{Friedmann}}}, \bauthor{\binits{H.O.}
  \bsnm{{Funsten}}}, \bauthor{\binits{A.G.} \bsnm{{Ghielmetti}}},
  \bauthor{\binits{J.} \bsnm{{Googins}}}, \bauthor{\binits{M.S.}
  \bsnm{{Granoff}}}, \bauthor{\binits{J.W.} \bsnm{{Hamilton}}},
  \bauthor{\binits{J.} \bsnm{{Hanley}}}, \bauthor{\binits{D.}
  \bsnm{{Heirtzler}}}, \bauthor{\binits{E.} \bsnm{{Hertzberg}}},
  \bauthor{\binits{D.} \bsnm{{Isaac}}}, \bauthor{\binits{B.} \bsnm{{King}}},
  \bauthor{\binits{U.} \bsnm{{Knauss}}}, \bauthor{\binits{H.}
  \bsnm{{Kucharek}}}, \bauthor{\binits{F.} \bsnm{{Kudirka}}},
  \bauthor{\binits{S.} \bsnm{{Livi}}}, \bauthor{\binits{J.} \bsnm{{Lobell}}},
  \bauthor{\binits{S.} \bsnm{{Longworth}}}, \bauthor{\binits{K.}
  \bsnm{{Mashburn}}}, \bauthor{\binits{D.J.} \bsnm{{McComas}}},
  \bauthor{\binits{E.} \bsnm{{M{\"o}bius}}}, \bauthor{\binits{A.S.}
  \bsnm{{Moore}}}, \bauthor{\binits{T.E.} \bsnm{{Moore}}},
  \bauthor{\binits{R.J.} \bsnm{{Nemanich}}}, \bauthor{\binits{J.}
  \bsnm{{Nolin}}}, \bauthor{\binits{M.} \bsnm{{O'Neal}}}, \bauthor{\binits{D.}
  \bsnm{{Piazza}}}, \bauthor{\binits{L.} \bsnm{{Peterson}}},
  \bauthor{\binits{S.E.} \bsnm{{Pope}}}, \bauthor{\binits{P.}
  \bsnm{{Rosmarynowski}}}, \bauthor{\binits{L.A.} \bsnm{{Saul}}},
  \bauthor{\binits{J.R.} \bsnm{{Scherrer}}}, \bauthor{\binits{J.A.}
  \bsnm{{Scheer}}}, \bauthor{\binits{C.} \bsnm{{Schlemm}}},
  \bauthor{\binits{N.A.} \bsnm{{Schwadron}}}, \bauthor{\binits{C.}
  \bsnm{{Tillier}}}, \bauthor{\binits{S.} \bsnm{{Turco}}}, \bauthor{\binits{J.}
  \bsnm{{Tyler}}}, \bauthor{\binits{M.} \bsnm{{Vosbury}}}, \bauthor{\binits{M.}
  \bsnm{{Wieser}}}, \bauthor{\binits{P.} \bsnm{{Wurz}}}, \bauthor{\binits{S.}
  \bsnm{{Zaffke}}},
\batitle{{The IBEX-Lo Sensor}}.
\bjtitle{Space Science Reviews}
\bvolume{146},
\bfpage{117}--\blpage{147}
(\byear{2009}a)
\end{barticle}
\endbibitem

\bibitem[\protect\citeauthoryear{{Fuselier} et~al.}{2009b}]{Fuselier:2009sci}
\begin{barticle}
\bauthor{\binits{S.A.} \bsnm{{Fuselier}}}, \bauthor{\binits{F.}
  \bsnm{{Allegrini}}}, \bauthor{\binits{H.O.} \bsnm{{Funsten}}},
  \bauthor{\binits{A.G.} \bsnm{{Ghielmetti}}}, \bauthor{\binits{D.}
  \bsnm{{Heirtzler}}}, \bauthor{\binits{H.} \bsnm{{Kucharek}}},
  \bauthor{\binits{O.W.} \bsnm{{Lennartsson}}}, \bauthor{\binits{D.J.}
  \bsnm{{McComas}}}, \bauthor{\binits{E.} \bsnm{{M{\"o}bius}}},
  \bauthor{\binits{T.E.} \bsnm{{Moore}}}, \bauthor{\binits{S.M.}
  \bsnm{{Petrinec}}}, \bauthor{\binits{L.A.} \bsnm{{Saul}}},
  \bauthor{\binits{J.A.} \bsnm{{Scheer}}}, \bauthor{\binits{N.}
  \bsnm{{Schwadron}}}, \bauthor{\binits{P.} \bsnm{{Wurz}}},
\batitle{{Width and Variation of the ENA Flux Ribbon Observed by the
  Interstellar Boundary Explorer}}.
\bjtitle{Science}
\bvolume{326},
\bfpage{962}
(\byear{2009}b)
\end{barticle}
\endbibitem

\bibitem[\protect\citeauthoryear{{Gruntman}}{1993}]{Gruntman:1993}
\begin{barticle}
\bauthor{\binits{M.A.} \bsnm{{Gruntman}}},
\batitle{{A new technique for in situ measurement of the composition of neutral
  gas in interplanetary space}}.
\bjtitle{\planss}
\bvolume{41},
\bfpage{307}--\blpage{319}
(\byear{1993}).
doi:\doiurl{10.1016/0032-0633(93)90026-X}
\end{barticle}
\endbibitem

\bibitem[\protect\citeauthoryear{{Grzedzielski}
  et~al.}{2010}]{GrzedzielskiBzowski:2010ribbon}
\begin{barticle}
\bauthor{\binits{S.} \bsnm{{Grzedzielski}}}, \bauthor{\binits{M.}
  \bsnm{{Bzowski}}}, \bauthor{\binits{A.} \bsnm{{Czechowski}}},
  \bauthor{\binits{H.O.} \bsnm{{Funsten}}}, \bauthor{\binits{D.J.}
  \bsnm{{McComas}}}, \bauthor{\binits{N.A.} \bsnm{{Schwadron}}},
\batitle{{A Possible Generation Mechanism for the IBEX Ribbon from Outside the
  Heliosphere}}.
\bjtitle{\apjl}
\bvolume{715},
\bfpage{84}--\blpage{87}
(\byear{2010})
\end{barticle}
\endbibitem

\bibitem[\protect\citeauthoryear{{Heerikhuisen}
  et~al.}{2010}]{Heerikhuisen:2010ribbon}
\begin{barticle}
\bauthor{\binits{J.} \bsnm{{Heerikhuisen}}}, \bauthor{\binits{N.V.}
  \bsnm{{Pogorelov}}}, \bauthor{\binits{G.P.} \bsnm{{Zank}}},
  \bauthor{\binits{G.B.} \bsnm{{Crew}}}, \bauthor{\binits{P.C.}
  \bsnm{{Frisch}}}, \bauthor{\binits{H.O.} \bsnm{{Funsten}}},
  \bauthor{\binits{P.H.} \bsnm{{Janzen}}}, \bauthor{\binits{D.J.}
  \bsnm{{McComas}}}, \bauthor{\binits{D.B.} \bsnm{{Reisenfeld}}},
  \bauthor{\binits{N.A.} \bsnm{{Schwadron}}},
\batitle{{Pick-Up Ions in the Outer Heliosheath: A Possible Mechanism for the
  Interstellar Boundary EXplorer Ribbon}}.
\bjtitle{\apjl}
\bvolume{708},
\bfpage{126}--\blpage{130}
(\byear{2010})
\end{barticle}
\endbibitem

\bibitem[\protect\citeauthoryear{{Hilchenbach}
  et~al.}{1998}]{Hilchenbachetal:1998ena}
\begin{barticle}
\bauthor{\binits{M.} \bsnm{{Hilchenbach}}}, \bauthor{\binits{K.C.}
  \bsnm{{Hsieh}}}, \bauthor{\binits{D.} \bsnm{{Hovestadt}}},
  \bauthor{\binits{B.} \bsnm{{Klecker}}}, \bauthor{\binits{H.}
  \bsnm{{Gruenwaldt}}}, \bauthor{\binits{P.} \bsnm{{Bochsler}}},
  \bauthor{\binits{F.M.} \bsnm{{Ipavich}}}, \bauthor{\binits{A.}
  \bsnm{{Buergi}}}, \bauthor{\binits{E.} \bsnm{{Moebius}}},
  \bauthor{\binits{F.} \bsnm{{Gliem}}}, \bauthor{\binits{W.I.}
  \bsnm{{Axford}}}, \bauthor{\binits{H.} \bsnm{{Balsiger}}},
  \bauthor{\binits{W.} \bsnm{{Bornemann}}}, \bauthor{\binits{M.A.}
  \bsnm{{Coplan}}}, \bauthor{\binits{A.B.} \bsnm{{Galvin}}},
  \bauthor{\binits{J.} \bsnm{{Geiss}}}, \bauthor{\binits{G.}
  \bsnm{{Gloeckler}}}, \bauthor{\binits{S.} \bsnm{{Hefti}}},
  \bauthor{\binits{D.L.} \bsnm{{Judge}}}, \bauthor{\binits{R.}
  \bsnm{{Kallenbach}}}, \bauthor{\binits{P.} \bsnm{{Laeverenz}}},
  \bauthor{\binits{M.A.} \bsnm{{Lee}}}, \bauthor{\binits{S.} \bsnm{{Livi}}},
  \bauthor{\binits{G.G.} \bsnm{{Managadze}}}, \bauthor{\binits{E.}
  \bsnm{{Marsch}}}, \bauthor{\binits{M.} \bsnm{{Neugebauer}}},
  \bauthor{\binits{H.S.} \bsnm{{Ogawa}}}, \bauthor{\binits{K.}
  \bsnm{{Reiche}}}, \bauthor{\binits{M.} \bsnm{{Scholer}}},
  \bauthor{\binits{M.I.} \bsnm{{Verigin}}}, \bauthor{\binits{B.}
  \bsnm{{Wilken}}}, \bauthor{\binits{P.} \bsnm{{Wurz}}},
\batitle{{Detection of 55-80 keV Hydrogen Atoms of Heliospheric Origin by
  CELIAS/HSTOF on SOHO}}.
\bjtitle{\apj}
\bvolume{503},
\bfpage{916}
(\byear{1998})
\end{barticle}
\endbibitem

\bibitem[\protect\citeauthoryear{{Hsieh} and
  {Gruntman}}{1993}]{HsiehGruntman:1993}
\begin{barticle}
\bauthor{\binits{K.C.} \bsnm{{Hsieh}}}, \bauthor{\binits{M.A.}
  \bsnm{{Gruntman}}},
\batitle{{Viewing the outer heliosphere in energetic neutral atoms}}.
\bjtitle{Advances in Space Research}
\bvolume{13},
\bfpage{131}--\blpage{139}
(\byear{1993}).
doi:\doiurl{10.1016/0273-1177(93)90402-W}
\end{barticle}
\endbibitem

\bibitem[\protect\citeauthoryear{{Hsieh} et~al.}{2010}]{Hsiehetal:2010ena}
\begin{barticle}
\bauthor{\binits{K.C.} \bsnm{{Hsieh}}}, \bauthor{\binits{J.}
  \bsnm{{Giacalone}}}, \bauthor{\binits{A.} \bsnm{{Czechowski}}},
  \bauthor{\binits{M.} \bsnm{{Hilchenbach}}}, \bauthor{\binits{S.}
  \bsnm{{Grzedzielski}}}, \bauthor{\binits{J.} \bsnm{{Kota}}},
\batitle{{Thickness of the Heliosheath, Return of the Pick-up Ions, and Voyager
  1's Crossing the Heliopause}}.
\bjtitle{\apjl}
\bvolume{718},
\bfpage{185}--\blpage{188}
(\byear{2010})
\end{barticle}
\endbibitem

\bibitem[\protect\citeauthoryear{{Krimigis} et~al.}{2009}]{Krimigis:2009sci}
\begin{barticle}
\bauthor{\binits{S.M.} \bsnm{{Krimigis}}}, \bauthor{\binits{D.G.}
  \bsnm{{Mitchell}}}, \bauthor{\binits{E.C.} \bsnm{{Roelof}}},
  \bauthor{\binits{K.C.} \bsnm{{Hsieh}}}, \bauthor{\binits{D.J.}
  \bsnm{{McComas}}},
\batitle{{Imaging the Interaction of the Heliosphere with the Interstellar
  Medium from Saturn with Cassini}}.
\bjtitle{Science}
\bvolume{326},
\bfpage{971}
(\byear{2009})
\end{barticle}
\endbibitem

\bibitem[\protect\citeauthoryear{{Kurth} and
  {Gurnett}}{2003}]{KurthGurnett:2003}
\begin{barticle}
\bauthor{\binits{W.S.} \bsnm{{Kurth}}}, \bauthor{\binits{D.A.}
  \bsnm{{Gurnett}}},
\batitle{{On the source location of low-frequency heliospheric radio
  emissions}}.
\bjtitle{\jgr}
\bvolume{108},
\bfpage{2}--\blpage{16}
(\byear{2003})
\end{barticle}
\endbibitem

\bibitem[\protect\citeauthoryear{{Lallement} et~al.}{2005}]{Lallementetal:2005}
\begin{barticle}
\bauthor{\binits{R.} \bsnm{{Lallement}}}, \bauthor{\binits{E.} \bsnm{{Qu{\'
  e}merais}}}, \bauthor{\binits{J.L.} \bsnm{{Bertaux}}}, \bauthor{\binits{S.}
  \bsnm{{Ferron}}}, \bauthor{\binits{D.} \bsnm{{Koutroumpa}}},
  \bauthor{\binits{R.} \bsnm{{Pellinen}}},
\batitle{{Deflection of the Interstellar Neutral Hydrogen Flow Across the
  Heliospheric Interface}}.
\bjtitle{Science}
\bvolume{307},
\bfpage{1447}--\blpage{1449}
(\byear{2005})
\end{barticle}
\endbibitem

\bibitem[\protect\citeauthoryear{{Lindsay} and
  {Stebbings}}{2005}]{LindsayStebbings:2005}
\begin{barticle}
\bauthor{\binits{B.G.} \bsnm{{Lindsay}}}, \bauthor{\binits{R.F.}
  \bsnm{{Stebbings}}},
\batitle{{Charge transfer cross sections for energetic neutral atom data
  analysis}}.
\bjtitle{Journal of Geophysical Research (Space Physics)}
\bvolume{110},
\bfpage{12213}
(\byear{2005})
\end{barticle}
\endbibitem

\bibitem[\protect\citeauthoryear{{McComas} et~al.}{2009a}]{McComas:2009sci}
\begin{barticle}
\bauthor{\binits{D.J.} \bsnm{{McComas}}}, \bauthor{\binits{F.}
  \bsnm{{Allegrini}}}, \bauthor{\binits{P.} \bsnm{{Bochsler}}},
  \bauthor{\binits{M.} \bsnm{{Bzowski}}}, \bauthor{\binits{E.R.}
  \bsnm{{Christian}}}, \bauthor{\binits{G.B.} \bsnm{{Crew}}},
  \bauthor{\binits{R.} \bsnm{{DeMajistre}}}, \bauthor{\binits{H.}
  \bsnm{{Fahr}}}, \bauthor{\binits{H.} \bsnm{{Fichtner}}},
  \bauthor{\binits{P.C.} \bsnm{{Frisch}}}, \bauthor{\binits{H.O.}
  \bsnm{{Funsten}}}, \bauthor{\binits{S.A.} \bsnm{{Fuselier}}},
  \bauthor{\binits{G.} \bsnm{{Gloeckler}}}, \bauthor{\binits{M.}
  \bsnm{{Gruntman}}}, \bauthor{\binits{J.} \bsnm{{Heerikhuisen}}},
  \bauthor{\binits{V.} \bsnm{{Izmodenov}}}, \bauthor{\binits{P.}
  \bsnm{{Janzen}}}, \bauthor{\binits{P.} \bsnm{{Knappenberger}}},
  \bauthor{\binits{S.} \bsnm{{Krimigis}}}, \bauthor{\binits{H.}
  \bsnm{{Kucharek}}}, \bauthor{\binits{M.} \bsnm{{Lee}}}, \bauthor{\binits{G.}
  \bsnm{{Livadiotis}}}, \bauthor{\binits{S.} \bsnm{{Livi}}},
  \bauthor{\binits{R.J.} \bsnm{{MacDowall}}}, \bauthor{\binits{D.}
  \bsnm{{Mitchell}}}, \bauthor{\binits{E.} \bsnm{{M{\"o}bius}}},
  \bauthor{\binits{T.} \bsnm{{Moore}}}, \bauthor{\binits{N.V.}
  \bsnm{{Pogorelov}}}, \bauthor{\binits{D.} \bsnm{{Reisenfeld}}},
  \bauthor{\binits{E.} \bsnm{{Roelof}}}, \bauthor{\binits{L.} \bsnm{{Saul}}},
  \bauthor{\binits{N.A.} \bsnm{{Schwadron}}}, \bauthor{\binits{P.W.}
  \bsnm{{Valek}}}, \bauthor{\binits{R.} \bsnm{{Vanderspek}}},
  \bauthor{\binits{P.} \bsnm{{Wurz}}}, \bauthor{\binits{G.P.} \bsnm{{Zank}}},
\batitle{{Global Observations of the Interstellar Interaction from the
  Interstellar Boundary Explorer (IBEX)}}.
\bjtitle{Science}
\bvolume{326},
\bfpage{959}
(\byear{2009}a)
\end{barticle}
\endbibitem

\bibitem[\protect\citeauthoryear{{McComas} et~al.}{2009b}]{McComasetal:2009ssr}
\begin{barticle}
\bauthor{\binits{D.J.} \bsnm{{McComas}}}, \bauthor{\binits{F.}
  \bsnm{{Allegrini}}}, \bauthor{\binits{P.} \bsnm{{Bochsler}}},
  \bauthor{\binits{M.} \bsnm{{Bzowski}}}, \bauthor{\binits{M.}
  \bsnm{{Collier}}}, \bauthor{\binits{H.} \bsnm{{Fahr}}}, \bauthor{\binits{H.}
  \bsnm{{Fichtner}}}, \bauthor{\binits{P.} \bsnm{{Frisch}}},
  \bauthor{\binits{H.O.} \bsnm{{Funsten}}}, \bauthor{\binits{S.A.}
  \bsnm{{Fuselier}}}, \bauthor{\binits{G.} \bsnm{{Gloeckler}}},
  \bauthor{\binits{M.} \bsnm{{Gruntman}}}, \bauthor{\binits{V.}
  \bsnm{{Izmodenov}}}, \bauthor{\binits{P.} \bsnm{{Knappenberger}}},
  \bauthor{\binits{M.} \bsnm{{Lee}}}, \bauthor{\binits{S.} \bsnm{{Livi}}},
  \bauthor{\binits{D.} \bsnm{{Mitchell}}}, \bauthor{\binits{E.}
  \bsnm{{M{\"o}bius}}}, \bauthor{\binits{T.} \bsnm{{Moore}}},
  \bauthor{\binits{S.} \bsnm{{Pope}}}, \bauthor{\binits{D.}
  \bsnm{{Reisenfeld}}}, \bauthor{\binits{E.} \bsnm{{Roelof}}},
  \bauthor{\binits{J.} \bsnm{{Scherrer}}}, \bauthor{\binits{N.}
  \bsnm{{Schwadron}}}, \bauthor{\binits{R.} \bsnm{{Tyler}}},
  \bauthor{\binits{M.} \bsnm{{Wieser}}}, \bauthor{\binits{M.} \bsnm{{Witte}}},
  \bauthor{\binits{P.} \bsnm{{Wurz}}}, \bauthor{\binits{G.} \bsnm{{Zank}}},
\batitle{{IBEX--Interstellar Boundary Explorer}}.
\bjtitle{Space Science Reviews}
\bvolume{146},
\bfpage{11}--\blpage{33}
(\byear{2009}b)
\end{barticle}
\endbibitem

\bibitem[\protect\citeauthoryear{{McComas} et~al.}{2010}]{McComasetal:2010var}
\begin{barticle}
\bauthor{\binits{D.J.} \bsnm{{McComas}}}, \bauthor{\binits{M.}
  \bsnm{{Bzowski}}}, \bauthor{\binits{P.C.} \bsnm{{Frisch}}},
  \bauthor{\binits{G.B.} \bsnm{{Crew}}}, \bauthor{\binits{M.A.}
  \bsnm{{Dayeh}}}, \bauthor{\binits{R.} \bsnm{{DeMajistre}}},
  \bauthor{\binits{H.O.} \bsnm{{Funsten}}}, \bauthor{\binits{S.A.}
  \bsnm{{Fuselier}}}, \bauthor{\binits{M.} \bsnm{{Gruntman}}},
  \bauthor{\binits{P.} \bsnm{{Janzen}}}, \bauthor{\binits{M.A.}
  \bsnm{{Kubiac}}}, \bauthor{\binits{G.} \bsnm{{Livadiotis}}},
  \bauthor{\binits{E.} \bsnm{{M{\"o}bius}}}, \bauthor{\binits{D.}
  \bsnm{{Reisenfeld}}}, \bauthor{\binits{N.A.} \bsnm{{Schwadron}}},
\batitle{"the evolving outer heliosphere: Large-scale stability and time
  variations observed by the interstellar boundary explorer"}.
\bjtitle{JGR}
\bvolume{00},
(\byear{2010})
\end{barticle}
\endbibitem

\bibitem[\protect\citeauthoryear{{M\"obius} et~al.}{1985}]{Moebiusetal:1985}
\begin{barticle}
\bauthor{\binits{E.} \bsnm{{M\"obius}}}, \bauthor{\binits{D.}
  \bsnm{{Hovestadt}}}, \bauthor{\binits{B.} \bsnm{{Klecker}}},
  \bauthor{\binits{M.} \bsnm{{Scholer}}}, \bauthor{\binits{G.}
  \bsnm{{Gloeckler}}},
\batitle{{Direct observation of He(+) pick-up ions of Interstellar Origin in
  the Solar Wind}}.
\bjtitle{\nat}
\bvolume{318},
\bfpage{426}--\blpage{429}
(\byear{1985})
\end{barticle}
\endbibitem

\bibitem[\protect\citeauthoryear{{M{\"o}bius} et~al.}{2004}]{Moebiusetal:2004}
\begin{barticle}
\bauthor{\binits{E.} \bsnm{{M{\"o}bius}}}, \bauthor{\binits{M.}
  \bsnm{{Bzowski}}}, \bauthor{\binits{S.} \bsnm{{Chalov}}},
  \bauthor{\binits{H.} \bsnm{{Fahr}}}, \bauthor{\binits{G.}
  \bsnm{{Gloeckler}}}, \bauthor{\binits{V.} \bsnm{{Izmodenov}}},
  \bauthor{\binits{R.} \bsnm{{Kallenbach}}}, \bauthor{\binits{R.}
  \bsnm{{Lallement}}}, \bauthor{\binits{D.} \bsnm{{McMullin}}},
  \bauthor{\binits{H.} \bsnm{{Noda}}}, \bauthor{\binits{M.} \bsnm{{Oka}}},
  \bauthor{\binits{A.} \bsnm{{Pauluhn}}}, \bauthor{\binits{J.}
  \bsnm{{Raymond}}}, \bauthor{\binits{D.} \bsnm{{Ruci{\'n}ski}}},
  \bauthor{\binits{R.} \bsnm{{Skoug}}}, \bauthor{\binits{T.}
  \bsnm{{Terasawa}}}, \bauthor{\binits{W.} \bsnm{{Thompson}}},
  \bauthor{\binits{J.} \bsnm{{Vallerga}}}, \bauthor{\binits{R.} \bsnm{{von
  Steiger}}}, \bauthor{\binits{M.} \bsnm{{Witte}}},
\batitle{{Synopsis of the interstellar He parameters from combined neutral gas,
  pickup ion and UV scattering observations and related consequences}}.
\bjtitle{\aap}
\bvolume{426},
\bfpage{897}--\blpage{907}
(\byear{2004})
\end{barticle}
\endbibitem

\bibitem[\protect\citeauthoryear{{M{\"o}bius} et~al.}{2009}]{Moebius:2009sci}
\begin{barticle}
\bauthor{\binits{E.} \bsnm{{M{\"o}bius}}}, \bauthor{\binits{P.}
  \bsnm{{Bochsler}}}, \bauthor{\binits{M.} \bsnm{{Bzowski}}},
  \bauthor{\binits{G.B.} \bsnm{{Crew}}}, \bauthor{\binits{H.O.}
  \bsnm{{Funsten}}}, \bauthor{\binits{S.A.} \bsnm{{Fuselier}}},
  \bauthor{\binits{A.} \bsnm{{Ghielmetti}}}, \bauthor{\binits{D.}
  \bsnm{{Heirtzler}}}, \bauthor{\binits{V.V.} \bsnm{{Izmodenov}}},
  \bauthor{\binits{M.} \bsnm{{Kubiak}}}, \bauthor{\binits{H.}
  \bsnm{{Kucharek}}}, \bauthor{\binits{M.A.} \bsnm{{Lee}}},
  \bauthor{\binits{T.} \bsnm{{Leonard}}}, \bauthor{\binits{D.J.}
  \bsnm{{McComas}}}, \bauthor{\binits{L.} \bsnm{{Petersen}}},
  \bauthor{\binits{L.} \bsnm{{Saul}}}, \bauthor{\binits{J.A.} \bsnm{{Scheer}}},
  \bauthor{\binits{N.} \bsnm{{Schwadron}}}, \bauthor{\binits{M.}
  \bsnm{{Witte}}}, \bauthor{\binits{P.} \bsnm{{Wurz}}},
\batitle{{Direct Observations of Interstellar H, He, and O by the Interstellar
  Boundary Explorer}}.
\bjtitle{Science}
\bvolume{326},
\bfpage{969}
(\byear{2009})
\end{barticle}
\endbibitem

\bibitem[\protect\citeauthoryear{{Opher} et~al.}{2009}]{OpherRichardson:2009}
\begin{barticle}
\bauthor{\binits{M.} \bsnm{{Opher}}}, \bauthor{\binits{J.D.}
  \bsnm{{Richardson}}}, \bauthor{\binits{G.} \bsnm{{Toth}}},
  \bauthor{\binits{T.I.} \bsnm{{Gombosi}}},
\batitle{{Confronting Observations and Modeling: The Role of the Interstellar
  Magnetic Field in Voyager 1 and 2 Asymmetries}}.
\bjtitle{Space Science Reviews}
\bvolume{143},
\bfpage{43}--\blpage{55}
(\byear{2009})
\end{barticle}
\endbibitem

\bibitem[\protect\citeauthoryear{{Pogorelov}
  et~al.}{2009}]{Pogorelovetal:2009L}
\begin{barticle}
\bauthor{\binits{N.V.} \bsnm{{Pogorelov}}}, \bauthor{\binits{J.}
  \bsnm{{Heerikhuisen}}}, \bauthor{\binits{J.J.} \bsnm{{Mitchell}}},
  \bauthor{\binits{I.H.} \bsnm{{Cairns}}}, \bauthor{\binits{G.P.}
  \bsnm{{Zank}}},
\batitle{{Heliospheric Asymmetries and 2-3 kHz Radio Emission Under Strong
  Interstellar Magnetic Field Conditions}}.
\bjtitle{\apjl}
\bvolume{695},
\bfpage{31}--\blpage{34}
(\byear{2009})
\end{barticle}
\endbibitem

\bibitem[\protect\citeauthoryear{{Rucinski} et~al.}{1996}]{Rucinskietal:1996}
\begin{barticle}
\bauthor{\binits{D.} \bsnm{{Rucinski}}}, \bauthor{\binits{A.C.}
  \bsnm{{Cummings}}}, \bauthor{\binits{G.} \bsnm{{Gloeckler}}},
  \bauthor{\binits{A.J.} \bsnm{{Lazarus}}}, \bauthor{\binits{E.}
  \bsnm{{Mobius}}}, \bauthor{\binits{M.} \bsnm{{Witte}}},
\batitle{{Ionization Processes in the Heliosphere - Rates and Methods of their
  Determination}}.
\bjtitle{\ssr}
\bvolume{78},
\bfpage{73}--\blpage{84}
(\byear{1996})
\end{barticle}
\endbibitem

\bibitem[\protect\citeauthoryear{{Schwadron} et~al.}{2009}]{Schwadron:2009sci}
\begin{barticle}
\bauthor{\binits{N.A.} \bsnm{{Schwadron}}}, \bauthor{\binits{M.}
  \bsnm{{Bzowski}}}, \bauthor{\binits{G.B.} \bsnm{{Crew}}},
  \bauthor{\binits{M.} \bsnm{{Gruntman}}}, \bauthor{\binits{H.} \bsnm{{Fahr}}},
  \bauthor{\binits{H.} \bsnm{{Fichtner}}}, \bauthor{\binits{P.C.}
  \bsnm{{Frisch}}}, \bauthor{\binits{H.O.} \bsnm{{Funsten}}},
  \bauthor{\binits{S.} \bsnm{{Fuselier}}}, \bauthor{\binits{J.}
  \bsnm{{Heerikhuisen}}}, \bauthor{\binits{V.} \bsnm{{Izmodenov}}},
  \bauthor{\binits{H.} \bsnm{{Kucharek}}}, \bauthor{\binits{M.} \bsnm{{Lee}}},
  \bauthor{\binits{G.} \bsnm{{Livadiotis}}}, \bauthor{\binits{D.J.}
  \bsnm{{McComas}}}, \bauthor{\binits{E.} \bsnm{{Moebius}}},
  \bauthor{\binits{T.} \bsnm{{Moore}}}, \bauthor{\binits{J.}
  \bsnm{{Mukherjee}}}, \bauthor{\binits{N.V.} \bsnm{{Pogorelov}}},
  \bauthor{\binits{C.} \bsnm{{Prested}}}, \bauthor{\binits{D.}
  \bsnm{{Reisenfeld}}}, \bauthor{\binits{E.} \bsnm{{Roelof}}},
  \bauthor{\binits{G.P.} \bsnm{{Zank}}},
\batitle{{Comparison of Interstellar Boundary Explorer Observations with 3D
  Global Heliospheric Models}}.
\bjtitle{Science}
\bvolume{326},
\bfpage{966}
(\byear{2009})
\end{barticle}
\endbibitem

\bibitem[\protect\citeauthoryear{{Slavin} and
  {Frisch}}{2008}]{SlavinFrisch:2008}
\begin{barticle}
\bauthor{\binits{J.D.} \bsnm{{Slavin}}}, \bauthor{\binits{P.C.}
  \bsnm{{Frisch}}},
\batitle{{The boundary conditions of the heliosphere: photoionization models
  constrained by interstellar and in situ data}}.
\bjtitle{\aap}
\bvolume{491},
\bfpage{53}--\blpage{68}
(\byear{2008})
\end{barticle}
\endbibitem

\bibitem[\protect\citeauthoryear{{Thomas} and
  {Krassa}}{1971}]{ThomasKrassa:1971}
\begin{barticle}
\bauthor{\binits{G.E.} \bsnm{{Thomas}}}, \bauthor{\binits{R.F.}
  \bsnm{{Krassa}}},
\batitle{{OGO 5 Measurements of the Lyman Alpha Sky Background}}.
\bjtitle{\aap}
\bvolume{11},
\bfpage{218}
(\byear{1971})
\end{barticle}
\endbibitem

\bibitem[\protect\citeauthoryear{{Vasyliunas}}{1968}]{Vasyliunas:1968}
\begin{barticle}
\bauthor{\binits{V.M.} \bsnm{{Vasyliunas}}},
\batitle{{A Survey of Low-Energy Electrons in the Evening Sector of the
  Magnetosphere with OGO 1 and OGO 3}}.
\bjtitle{\jgr}
\bvolume{73},
\bfpage{2839}
(\byear{1968})
\end{barticle}
\endbibitem

\bibitem[\protect\citeauthoryear{{Weller} and {Meier}}{1974}]{WellerMeier:1974}
\begin{barticle}
\bauthor{\binits{C.S.} \bsnm{{Weller}}}, \bauthor{\binits{R.R.}
  \bsnm{{Meier}}},
\batitle{Observations of helium in the interplanetary/interstellar wind - the
  solar-wake effect}.
\bjtitle{\apj}
\bvolume{193},
\bfpage{471}--\blpage{476}
(\byear{1974})
\end{barticle}
\endbibitem

\end{thebibliography}

\end{document}